\documentclass[]{article}
\usepackage{graphicx,multirow}

\begin{document}

\title{Shadows of the SIS immortality transition in small networks}
\author{Petter Holme%
\thanks{Department of Energy Science, Sungkyunkwan University, 440--746 Suwon, Republic of Korea}}

\begin{abstract}
Much of the research on the behavior of the SIS model on networks has concerned the infinite size limit; in particular the phase transition between a state where outbreaks can reach a finite fraction of the population, and a state where only a finite number would be infected. For finite networks, there is also a dynamic transition---the immortality transition---when the per-contact transmission probability $\lambda$ reaches one. If $\lambda < 1$, the probability that an outbreak will survive by an observation time $t$ tends to zero as $t \rightarrow \infty$; if $\lambda = 1$, this probability is one. We show that treating $\lambda = 1$ as a critical point predicts the $\lambda$-dependence of the survival probability also for more moderate $\lambda$-values. The exponent, however, depends on the underlying network. This fact could, by measuring how a vertex' deletion changes the exponent, be used to evaluate the role of a vertex in the outbreak. Our work also confirms an extremely clear separation between the early die-off (from the outbreak failing to take hold in the population) and the later extinctions (corresponding to rare stochastic events of several consecutive transmission events failing to occur).
\end{abstract}


\maketitle

\section{Introduction}

The Susceptible-Infectious-Susceptible model captures the dynamics of an infectious disease spreading in a population where infected people, upon recovery, become susceptible again. It has a long history, both as practical tool for predicting and understanding real outbreaks (e.g.\ Ref.~\cite{hy}), and as a problem in applied mathematics (as a special case of stochastic logistic processes)~\cite{no,kl,naa,do1,do2,o2,pvm,cps,bl,bo,ps}, statistical physics and computational science~\cite{pvm,cps,bl,bo,ps,go}. The theoretical development has, for example, focused on fully connected topologies (or well-mixed models in epidemiological jargon) where every individual is equally likely to meet everyone else, at every unit of time. One of the recent advances is to calculate the exact value of the average extinction time in a finite system~\cite{no,kl,naa}. In parallel, one of the greatest advances of theoretical epidemiology the last two decades is to move away from the well-mixed assumption and model the contact over which the disease spreads as a network~\cite{ps,mo,kee}. In this direction, researchers have, for example, calculated the epidemic threshold for a given network~\cite{pvm,cps,bo,ps}---i.e.\ the critical value of the per-contact transmission probability $\lambda$, below which a disease cannot reach a finite fraction of the population in the $N \rightarrow\infty$ limit.

For finite sized networks, there is also a phase transition, at least a threshold phenomenon albeit one whose value is trivial. If $\lambda < 1$, any outbreak will eventually die out in a finite network; if $\lambda = 1$, they will live forever. This immortality transition may be an epidemiological curiosity, as assumptions, such that the underlying network is relatively static, would break down for large enough $\lambda$-values. It is also trivial in the sense that it is not an emergent phenomenon. Still, it could be the case that the increasingly unlikely extinction events as $\lambda$ grows follow the same statistics as the critical slowing down around thermodynamic phase transition. In other words, we cannot \textit{a priori} exclude the possibility of something like a critical behavior of the immortality transition. So, since we cannot rule it out, we will assume that it is indeed true and see where it leads us.

In the remainder of this paper, we will investigate the scaling of the \textit{survival probability} $\xi$---the chance, as a function of $\lambda$, an outbreak survives past an observation time $t$. Furthermore, we will investigate the contribution of individual vertices to the behavior of $\xi$. This, we will argue, gives a new way of looking at importance of vertices in the SIS model on networks.

\section{Preliminaries}

\begin{figure}
  \includegraphics[width=0.9\columnwidth]{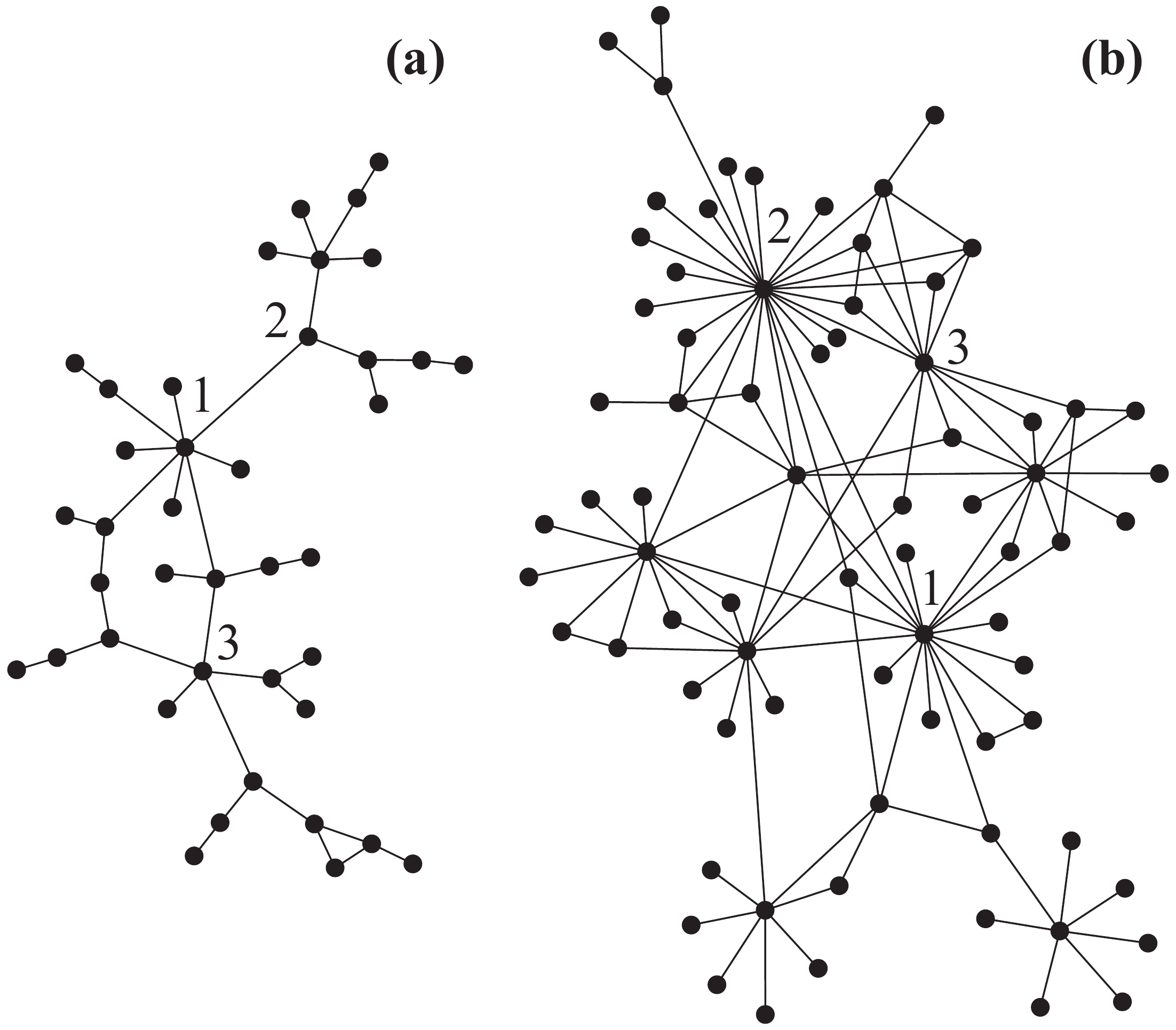}
  \caption{The two networks that we study (with the tree vertices of largest $a_i$-values marked). Panel (a) shows a sexual network from the beginning of the American HIV outbreak of the late 1970's from Ref.~\cite{au}; (b) shows a sexual network of Icelandic MSM from Ref.~\cite{ha}.
}
\label{f1}
\end{figure}

We use the constant-duration version of the SIS model defined as follows. Assume an underlying network represented as a simple graph $G = (V,E)$, where $V$ is a set of $N$ vertices and $E$ is a set of $M$ edges. At the beginning of the simulation, all vertices are susceptible except one vertex that is infectious (or rather, becomes infectious at this very time step). Then, for every edge between an infectious and a susceptible, the susceptible will become infectious with a probability $\lambda$ the following time step. An infectious vertex becomes susceptible again $\delta$ time steps after becoming infectious. When there are no infectious vertices, the outbreak is dead. For large enough, $\delta$ this is effectively a one-parameter model---the product $\lambda\delta$ determines the entire behavior of the model, their actual values only set the time scale. If $\delta$ is smaller than the shortest time for an outbreak to spread through the graph, there might be other effects arising. Ref.~\cite{ho} discusses this issue and argues that for practical purposes the constant-duration version of compartmental models is equivalent to the constant-recovery rate version (the latter being more common in the mathematical literature). (But a disclaimer is that Ref.~\cite{ho} did not deal with extreme events such as extinctions.) We use $\delta = 5$, which is around the radius of the graphs we study (and thus of the order of the fastest times of the disease to spread across the network).

Our core quantity for monitoring the outbreaks is the survival probability $f(\lambda,t)$---the fraction of outbreaks that are dead by time $t$. $df/dt$ gives probability distribution of extinction times, which is perhaps a more a commonly studied property~\cite{no,kl,do1,do2,o2}. All simulations are averaged over $10^6$ independent simulation runs.

For the purpose of this paper, any kind of small and somewhat heterogeneous network would suffice. But since there are such empirical networks in the epidemiology literature, we will take two of them as our study objects. Both networks represent sexual networks of men who have sex with men (MSM) and were collected to study the HIV epidemics. Of course, other compartmental models are more appropriate than SIS for modeling HIV~\cite{ko}, but there are other infections that spread over these networks---like chlamydia, syphilis and gonorrhea---that fits the SIS picture. Since these infections have different transmission pathways among MSM, epidemiologists typically treat MSM as a special case; also for diseases other than HIV~\cite{be}.

The first example network is based on contact tracing (following infection chains backward in time) in the early HIV outbreak in Canada and the United States~\cite{au} (we refer to this data as \textit{America}). It has $N = 40$ and $M = 41$. Note that contact tracing induces structures in the sampled networks that do not have to be present in the underlying sexual networks~\cite{ek}. Then again, this networks serves well as an example. Another network represent the MSM network of Iceland in the early 1990's and probably suffers less from sampling bias than the previous network~\cite{ha}. Here, $N = 75$ and $M = 115$. We refer to this data \textit{Iceland}. Both networks are visualized in Fig.~\ref{f1}.

\begin{figure}
  \includegraphics[width=\columnwidth]{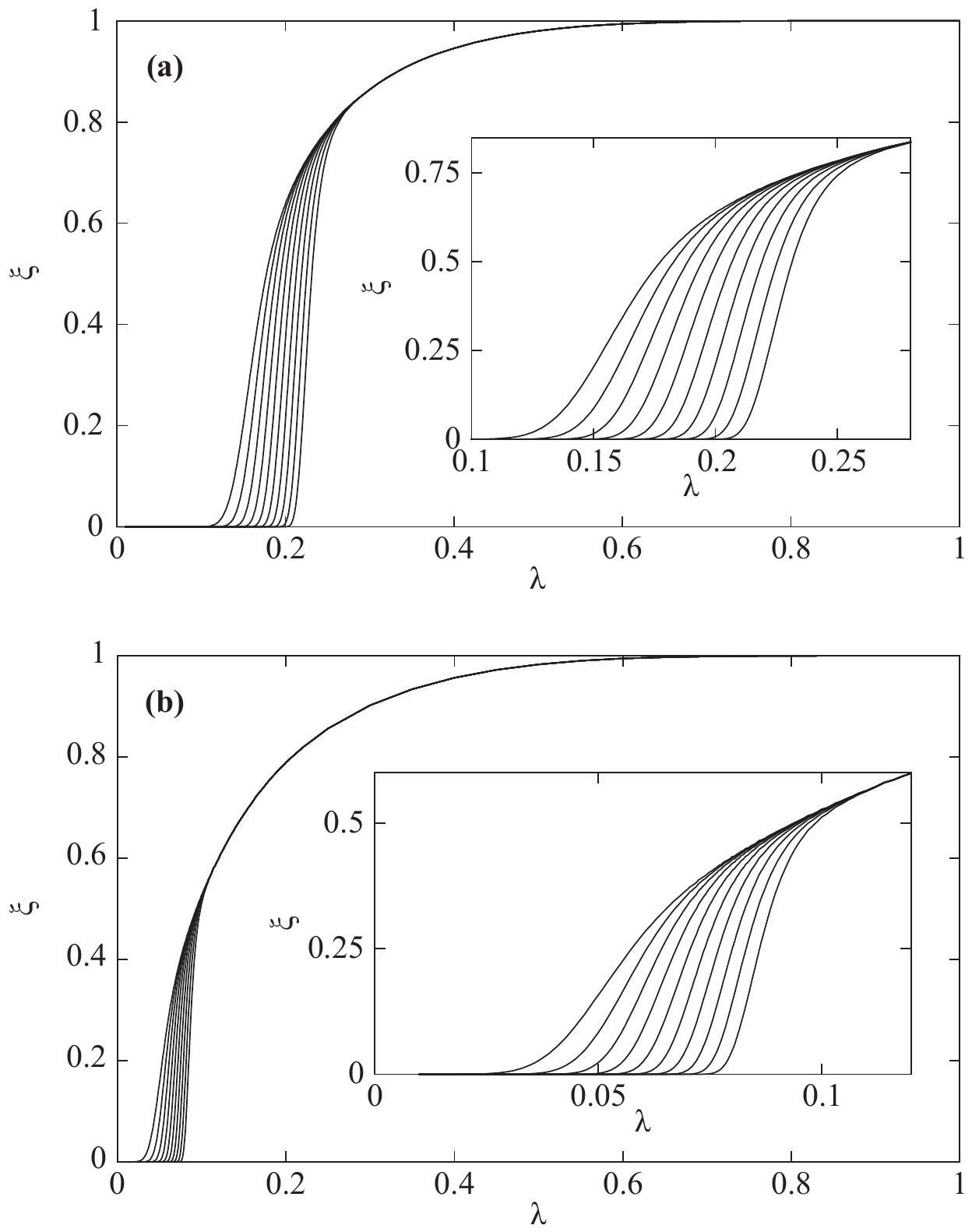}
  \caption{In panels (a) (for the \textit{America} data) and (b) (for the \textit{Iceland} data), we see the survival probability $\xi$ as a function of $\lambda$ for an exponentially growing set of observation times $t = 200\times 2^\nu$   in (a) and $t = 50\times 2^\nu$  in (b), where $\nu = 0,\dots,9$ for both panels. The lower $t$ value a curve has, the more to the left it is.
}
\label{f2}
\end{figure}

\begin{figure}
  \includegraphics[width=\columnwidth]{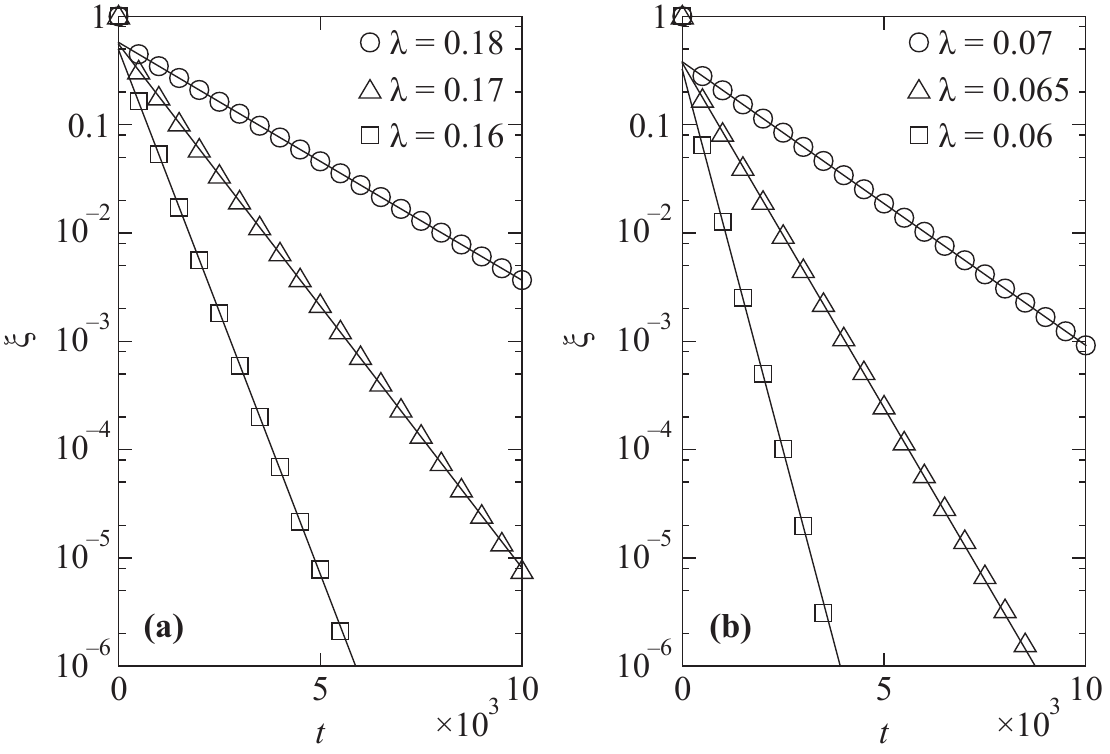}
  \caption{Fits of the extinction time data to an exponential functional form to determine the time constant $\tau$. Panel (a) shows graphs for \textit{America}; panel (b) is a corresponding plot for the \textit{Iceland}. Error bars are smaller than the symbol sizes.
}
\label{f3}
\end{figure}

\section{Results}

\subsection{The scaling of the survival probability}

We start our expose of numerical results by plotting the survival probability for fixed observation times $t$ as a function of $\lambda$ (Fig.~\ref{f2}). This type of plot is reminiscent of the average outbreak size $\Omega$ as a function of $\lambda$, which for infinite systems show a threshold phenomenon. $\Omega$ (precisely defined as the fraction of individuals that at some time are infected) changes, at a critical $\lambda$-value, from zero to $\Omega > 0$. At a first glance, it seems like the same thing happens in Fig.~\ref{f2}. At a certain $\lambda$ value, $\xi$ increases very rapidly from a value close to zero. The increase becomes steeper the larger the $t$-value is, so $t$ seems to take the role of $N$ in finite-size analysis of the phase transition in $\Omega$. As mentioned, we already know that there is a threshold behavior in this case, but at $\lambda = 1$. At a closer look, we can see that the rise of the $\xi$-curves happens later for every larger $t$-value. Since we use an exponential progression of $t$-values this increase is in effect very slow. Another conspicuous feature of these curves is that they all group into one for large enough $t$-values. This envelope of curves (or, rather, pseudo envelope, since it will disappear as $t \rightarrow \infty$) corresponds to the early die-off seen in many epidemic models. More precisely, one minus the $\xi$ value of the envelope gives the fraction of outbreaks that fail to take hold in the population. The remaining runs do reach some quasi-endemic state, but will eventually die due to the fluctuations in a finite network.

\begin{figure}
  \includegraphics[width=\columnwidth]{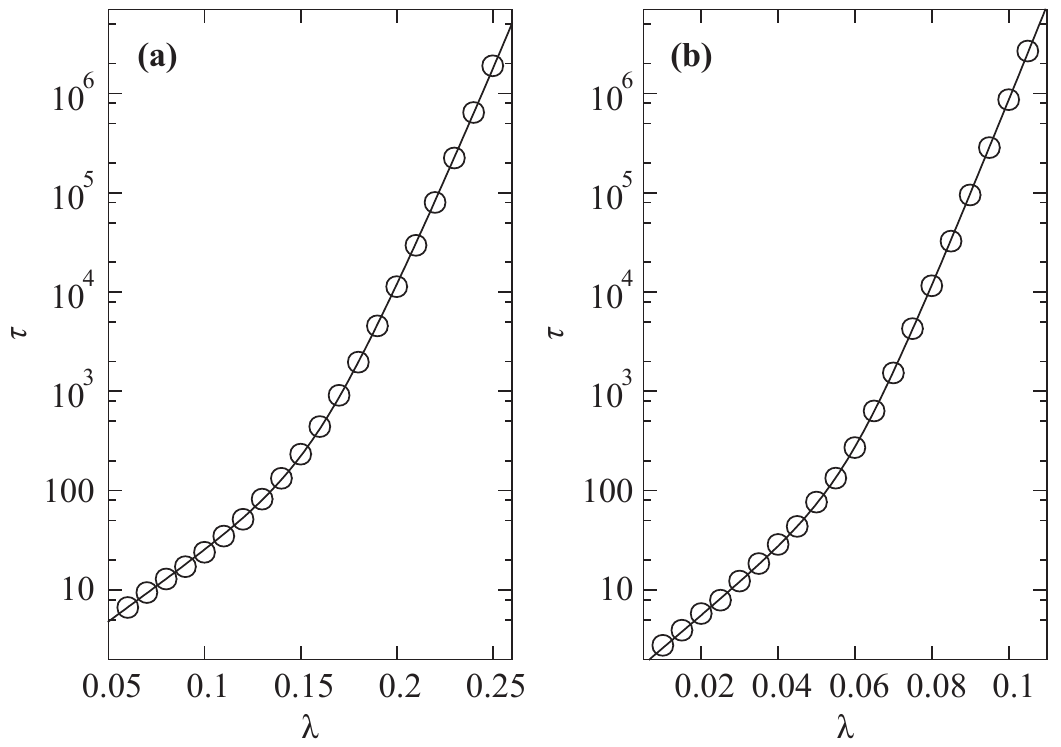}
  \caption{The time constant of the decay of $\xi$, $\tau$, as a function of $\lambda$. Panel (a) gives results for the \textit{America} data; panel (b) for the \textit{Iceland} data. The curves are fits to the form $A\exp(\lambda/l)+B(1-\lambda)^{-\zeta}$. The parameter values for the fit are $A=0.94(7)$, $\lambda=0.0307(8)$, $B=2.9(5)\times10^{-4}$ and $\zeta=78.6(6)$ for \textit{America}, and $A=1.22(2)$, $\lambda=0.0132(1)$, $B=8.2(4)\times 10^{-4}$ and $\zeta=197(6)$ for \textit{Iceland} (the digits in the parentheses are the standard errors in order of the last digits).
}
\label{f4}
\end{figure}

Next we look closer at $\xi$ as a function of $t$ for fixed $\lambda$-values. See Fig.~\ref{f3}. We chose $\lambda$-values to get a large range in $\xi$ (i.e.\ around where curves are the steepest in Fig.~\ref{f2}), but for both networks and all $\lambda$ the observation is the same---except very small $t$, $\xi(t)$ follows an exponential function $\exp(-t/\tau)$ quite accurately. This has been found analytically for fully connected networks~\cite{no,naa,o2}, and we guess that it holds for all connected networks. The exact functional form of the early die-off (where $\xi(t)$ decays faster than an exponential) could depend both on the heterogeneities of the networks around the seed and the discrete time. We will not go deeper into this, but focus on $\tau$. There are plenty of articles on the behavior of the average $\tau$ in fully connected networks in the large-$N$ limit~\cite{kl,do1,do2,o2}. Perhaps the theories in these references could be extended to the finite-$N$ and derive $\tau(\lambda)$. As mentioned, we take a  computational physics approach. Our next step is to construct a scaling ansatz for $\tau(\lambda)$. Assuming two independent scaling regimes where the large-$\lambda$ one is dominated by fluctuations similar to a critical point from the $\lambda = 1$ transition---$\tau \sim (1 - \lambda)^{-\zeta}$ where $\zeta$ is an exponent corresponding to a critical exponent. For the low-$\lambda$ part, the fully connected, $N \rightarrow \infty$ case predicts a linear dependence~\cite{naa}, but that does not fit our numerical results, which seem to be exponential. We combine (more precisely, add, assuming they stem from independent mechanisms) this observation with the finite-size scaling form to get
\begin{equation}\label{e1}
\tau=A\exp(\lambda/l)+B(1-\lambda)^{-\zeta}
\end{equation}
where $A$ and $B$ are constants, $l$ controls the small $\lambda$ behavior and $\zeta$ determines the  dynamics close to the immortality transition. Fitting to the form given by Eq.~(\ref{e1}) is very accurate (Fig.~\ref{f4}). Fits to other four-parameter functional forms (with a linear or power-law scaling of the low-$\lambda$ term), is visibly worse (failing at low $\lambda$). Unfortunately, only limited regions of $\lambda$ are accessible---for small $\lambda$, the disease dies too fast to get reliable data; for large $\lambda$, the simulations take too much time. The accuracy of the scaling form is probably higher than many numerical phase transition studies, which makes this somewhat more tolerable.  Our conclusion from Fig.~\ref{f4} is that it is consistent with a critical scaling around the $\lambda = 1$ transition affecting the dynamics at much lower $\lambda$. At the same time, we hope future studies could derive both terms of Eq.~(\ref{e1}) in a more systematic way. As a final note, the values of $\zeta$ are both high (compared to critical exponents) and different for the two networks. These both observations suggest that the network topology determines the exponent. This is different from the universality classes of spin systems on lattices where the exponents take rather few values.

\subsection{Vertex-vise contribution to $\zeta$}

\begin{table}
\begin{tabular}{l|l|l|l}
\multicolumn{2}{l|}{Measure} & \textit{America} & \textit{Iceland} \\\hline
\multirow{4}{*}{\rotatebox{90}{0-param.}} & $k_i$ & 0.73(4) & 0.974(2) \\
\multirow{4}{*}{} & $n_i$ & 0.82(4) & 0.75(5) \\
\multirow{4}{*}{} & $m_i$ & 0.83(3) & 0.965(2) \\
\multirow{4}{*}{} & $\epsilon_i$ & 0.64(4) & 0.917(6) \\\hline
\multirow{4}{*}{\rotatebox{90}{1-param.}} & $\max K_i$ & $0.76(5)$ & $0.98(2)$\\
\multirow{4}{*}{} & for $\alpha$ & $0.17(8)$ & $0.038(5)$\\
\multirow{4}{*}{} & $\max R_i$ & 0.72(6) & 0.97(4) \\
\multirow{4}{*}{} & for $d$ & 0.99(1) & 0.99(1)
\end{tabular}
\caption{
Pearson's correlation coefficient between $a_i$---capturing a vertex' influence on $\zeta$---and standard descriptors of the position of a vertex in a network. The number in the parenthesis is the standard error (obtained through jackknife resampling) in the order of the last digit. For the one-parameter quantities we also show the optimizing parameter values ($\alpha$ for Katz centrality and $d$ for PageRank).
}\label{t1}
\end{table}

We have seen that $\zeta$ depends on the network, but how? Now we turn to examining the contribution of each vertex to the scaling behavior of $\zeta$. We quantify the contribution of vertex $i$ by the ratio a between $\zeta$ after and before $i$ is removed. Let $G_i$ represent $G$ with i deleted, then:
\begin{equation}\label{e2}
a_i=\zeta(G_i)/\zeta(G)
\end{equation}
For all the vertices in our networks, $\zeta(G_i)$ is less than $\zeta(G)$, or within one standard error from it. It is easy to understand that adding a vertex is like making the road to extinction one step longer and thus even harder to reach, but how this translates to the more dramatic divergence for large $\lambda$ is not clear. We leave the observation that a larger graph has larger $\zeta$ as a conjecture and focus on the network structural predictors of $a_i$.

First, we investigate parameter-free descriptors of $i$'s position in the network, such as degree $k_i$, the number of vertices $n_i$ and edges $m_i$ in the largest connected component of $G_i$, coreness, closeness, betweenness, and the eigenvector centrality $\epsilon_i$. The latter comes from the idea that a vertex' centrality is, recursively, proportional to the sum of its neighbors' centralities. This leads to $\epsilon_i$ being the $i$'th element of $G$'s adjacency matrix (where the element on the $i$'th row and $j$'th column is one of $i$ and $j$ form an edge, and zero otherwise). All the quantities we use are discussed in network theory textbooks such as Ref.~\cite{ne}. We will consider the ones that have the highest explanatory power (measured through the absolute value of Pearson's correlation coefficient $r$) with respect to $a_i$, namely $k_i$ , $n_i$, $m_i$. and $\epsilon_i$. The $r$-values between these measures and a are presented in Table~\ref{t1}. Even though the quantities that we do not list (closeness, etc.) are consistently worse, the two data sets show quite different results for the listed quantities. Note, from Fig.~\ref{f1}, that the \textit{Iceland} network is still connected into one large component no matter which vertex that is deleted. In the \textit{America} data, on the other hand, deleting the vertices with the largest $a_i$ values disconnects the network. For this latter data set, the quantities that capture the fragmentation of the network, i.e.\ $n_i$ and $m_i$, are the best predictors of $a_i$. In Fig.~\ref{f1}, we plot the three vertices of largest $a_i$ for the two networks, which illustrates this point well.  For the other network, \textit{Iceland}, degree and $m_i$ (that in this case---with little fragmentation), is strongly correlated with degree, are the ones with highest correlation with $a_i$. Interestingly eigenvector centrality is performing poorly, highlighting the difference between indirect interaction in the SIS model and the linear coupling that the eigenvector centrality builds on. We note the ranges of $a_i$ is $[1.06(7),1.9(1)]$ for \textit{America} and $[1.004(2),1.24(5)]$ for \textit{Iceland} (where the numbers in parentheses are standard errors in the order of the last digit).

In addition to the zero-parameter importance measures above, we also measure the best possible correlation for two one-parameter measures---Katz centrality $K_i$ and PageRank $R_i$. These measures are related to the eigenvector centrality---Katz centrality also assume the centrality is proportional to the sum of the centrality of the neighbors, but also plus a constant $\alpha$ for the vertex itself. PageRank is proportional to the occupation probability of an unbiased random walker that with a probability $d$ moves to a neighbor of its current vertex, and otherwise it moves to a random vertex. The fact that a disease can spread from one vertex to many others make Katz centrality seem more appropriate. We include PageRank for comparison. The optimal correlation with Katz centralities is indeed always stronger than for the PageRank. The PageRank is optimized in the small-$d$ limit and the Katz centrality for an intermediate $\alpha$ value. For the \textit{Iceland} data the Katz centrality shows the highest correlation of all, meaning that there are measurable higher order structures that captures $a_i$ better than degree. For the \textit{America} data, $n_i$ and $m_i$ are still the measures with highest correlation with $a_i$, but after those come Katz centrality. In summary, for our somewhat sketchy analysis, how much deleting a vertex would fragment the network seems to be the most important structure for explaining $a_i$, degree is the second most important (but not the only other) factor.

\section{Discussion}

We have numerically investigated extinction events in the SIS model on small networks. Our observations are consistent with the assumption that the transition at $\lambda = 1$---below which an outbreak would always die out in a finite-sized network---can be treated with standard finite-size scaling theory, but with the size replaced by time (not an entirely new idea, cf.\ Ref.~\cite{mj}). Using this assumption, we find exponents that are dependent on the network topology. The fact that the $\zeta$ depend on the topology is different from critical phenomena where exponents are groped into universality classes. One interpretation is to see the network as an integral part of the model---after all, we we have already accepted to drop size-scaling from the picture. On the other hand, we could see this as an indication not to push the analogy between the immortality transition and critical behavior further.
We argue that $\zeta$ can be used as a parameter-free index of a vertex' role in the SIS extinction dynamics. We define the index as the exponent for the network without the vertex, divided by the exponent with the vertex present. This index, we show, depends much (but not only) on how deleting the vertex would fragment the network.

The extinction time for large $\lambda$ is extremely long. Even though our networks are small, if $\lambda > 1/2$, one would have to wait more than $10^{10^6}$) time steps for the survival fraction to go below 50\%, even in the smallest networks. In our analysis, a time step represents a fifth of the duration of the disease. This means that for diseases lasting a week or so, we would have to wait $10^5$ times the age of the universe (and many more times an average lifespan, let alone the typical lifetime of an edge in the network) to see more than half of the outbreak die out. In addition, the extinction times grow fast with $N$ (for fully connected networks, the growth of the average is exponential~\cite{naa}). From this discussion, we understand that no real outbreak would ever come close to the immortality transition without violating the assumption of a stable underlying network. Indeed, no direct SIS simulation comes close either.

For practical purposes, the most important observation is the extremely clear separation between the early die-off of outbreaks that fail to take hold in the population, and the later extinction events from rare stochastic events. This is maybe most clearly visible as the envelope of the curves in Fig.~\ref{f1}. This observation has been made before---e.g.\ Ref.~\cite{kl} argues for the relevance of the quasi-stationary stage of the SIS dynamics. Since awareness and countermeasures can, effectively, rewire the network fairly quickly only the early extinction events are of practical interest. The fraction of early extinctions is thus, from a modeling point of view, very well defined, which makes it easier to make conclusive statements from simulations. From a theoretical point of view, the crossover from exponential to power-law divergence scaling (Eq.~(\ref{e1})) is also interesting and opens for future investigations.

\section*{Acknowledgments}
The author acknowledges suggestions from Beom Jun Kim, Mark Newman, Jari Saram\"{a}ki, and Taro Takaguchi. The author was supported by Basic Science Research Program through the National Research Foundation of Korea (NRF) funded by the Ministry of Education (2013R1A1A2011947).

\end{document}